\documentclass[showpacs,aps,prx,twocolumn,amsmath,amssymb,epsfig]{revtex4-1}

\usepackage{graphicx}
\usepackage{epsfig}
\usepackage{dcolumn}
\usepackage{bm}
\newcommand     {\beq}[1]         { \begin{equation} #1 \end{equation} }

\usepackage{graphicx}
\usepackage{amssymb,amsfonts,amsmath}

\begin{document}


\title{Emergence of energy dependence in the fragmentation of heterogeneous materials}

\author{Gerg\H o P\'al$^1$, Imre Varga$^2$, and 
Ferenc Kun$^{1}$\email{ferenc.kun@science.unideb.hu}}

\affiliation{$^1$Department of Theoretical Physics, 
University of Debrecen, H-4010 Debrecen, P.O.Box: 5, Hungary}
\affiliation{$^2$Department of Informatics Systems and Networks, University of 
Debrecen,
P.O. Box 12, H-4010 Debrecen, Hungary}

\begin{abstract}
The most important characteristics of the fragmentation of heterogeneous 
solids is that the mass (size) distribution 
of pieces is described by a power law functional form. The exponent of the distribution 
displays a high degree of universality depending mainly on the dimensionality 
and on the brittle-ductile mechanical response of the system. Recently,
experiments and computer simulations have reported an energy dependence of
the exponent increasing with the imparted energy.
These novel findings question the phase transition picture 
of fragmentation phenomena, and have also practical importance for industrial applications.
Based on large scale computer simulations here we uncover a robust mechanism which leads to
the emergence of energy dependence in fragmentation processes 
resolving controversial issues on the problem:
studying the impact induced breakup of plate-like objects with varying thickness 
in three dimensions we show that energy dependence 
occurs when a lower dimensional fragmenting object 
is embedded into a higher dimensional space. The reason is an underlying transition 
between two distinct fragmentation mechanisms controlled by the impact velocity 
at low plate thicknesses, while it is hindered for three-dimensional bulk systems.
The mass distributions of the subsets of fragments dominated by the two cracking 
mechanisms proved to have an astonishing robustness at all plate thicknesses, which
implies that the non-universality of the complete mass distribution is the consequence 
of blending the contributions of universal partial processes.
\end{abstract}

\maketitle

\section{Introduction}
Fragmentation into numerous pieces occurs when a large
amount of energy is imparted to a solid within a short time 
\cite{astrom_statistical_2006}. 
Impact induced 
fragmentation of heterogeneous materials is abundant in nature having also 
a high importance for industrial applications especially in mining and ore processing 
\cite{stronge_impact_book,hayakawa_impact_1996,katsuragi_scaling_2003,
katsuragi_crossover_2004,katsuragi_explosive_2005,dos_santos_experimental_2010,
ji_fragmentation_2010}.
During the past decades research on fragmentation mainly focused
on the statistics of fragment masses (sizes) obtained by the breakup
of heterogeneous materials 
\cite{turcotte_factals_1986,turcotte_fractals_1997,astrom_statistical_2006}. 
A large number of experimental \cite{oddershede_self-organized_1993,inaoka_aspect_1997,astrom_statistical_2006,
turcotte_fractals_1997,turcotte_factals_1986,
astrom_universal_2004,wittel_fragmentation_2004,katsuragi_crossover_2004,katsuragi_explosive_2005,
kun_scaling_2006,dos_santos_experimental_2010,ji_fragmentation_2010,PhysRevE.84.026115,hooper_impact_2012}
and theoretical studies \cite{kun_transition_1999,astrom_universal_2004,behera_fragmentation_2005,
dos_santos_schematic_2007,
carmona_fragmentation_2008,
durand_jap_2012,PhysRevE.86.016113,Myagkov2014120} 
have confirmed that the mass distribution of 
fragments is described by a power law functional form.
The exponent of the distribution was found to show a high degree of
robustness, i.e.\  investigations revealed that the value of 
the exponent does not depend on the type of materials, amount of input 
energy and on the way the energy is imparted to the system until materials 
of a high degree of heterogeneity are fragmented 
\cite{astrom_statistical_2006,turcotte_fractals_1997,turcotte_factals_1986}. The value of the exponent
is mainly determined by the dimensionality of the system 
\cite{kun_study_1996,kun_transition_1999,wittel_fragmentation_2004-1,
behera_fragmentation_2005,kun_scaling_2006,carmona_fragmentation_2008,
PhysRevE.86.016113,astrom_universality_2000,astrom_universal_2004} 
and by the brittle or ductile mechanical response of the material 
\cite{timar_new_2010}. The universality of fragmenting has been shown to be the fingerprint
of an underlying phase transition from the damaged to the fragmented phase of the breakup
process \cite{kun_transition_1999,katsuragi_crossover_2004,katsuragi_explosive_2005,wittel_fragmentation_2004}.

Recently, experiments on the impact induced fragmentation of long thin glass rods 
\cite{ishii_fragmentation_1992,Ching200083}
and freely-hanging glass plates \cite{kadono_crack_2002,kadono_fragment_2005} 
revealed energy dependence of the mass distribution 
exponent, i.e.\ the exponent was found to increase with the imparted 
energy 
\cite{ishii_fragmentation_1992,Ching200083,kadono_crack_2002,kadono_fragment_2005,
myagkov_critical_2005,sator_damage_2010,n._sator_generic_2008}. 
The importance of these novel findings originates from the fact that, on the one 
hand, they question the universality and hence the phase transition interpretation 
of fragmentation phenomena, and on the other hand, they have consequences on the 
design of engineering technologies used for crushing in mining and ore processing 
\cite{turcotte_fractals_1997}.
Recent computer simulations have also provided an interesting counter example 
\cite{PhysRevE.86.016113}:
mass distributions of pieces obtained by the breakup of spherical bodies 
impacting against a hard wall have been found to get steeper with increasing 
impact velocity.
However, it proved to be an apparent energy dependence which occurs solely due to
the moving cutoff of the mass distributions, and hence, it can be transformed out
by rescaling with the average fragment mass \cite{PhysRevE.86.016113}. This study 
highlighted the importance of scaling and data collapse analysis when evaluating 
fragmentation results of finite size systems.

In order to resolve controversial issues on the energy dependence of the 
exponent of fragment mass distributions, in the present paper we
study the impact induced breakup of heterogeneous materials by large scale 
computer
simulations. Our results demonstrate that energy dependence emerges 
when the fragmenting object is embedded in a higher dimensional space.
Studying the fragmentation of plate-like objects in 
three dimensions (3D) we show that energy dependence is obtained 
for low plate thicknesses, while
it disappears for thick plates. The reason is that due to the interplay of the 
geometry of the sample and of the embedding space a transition 
takes place in the system 
between two fragmentation mechanisms as the impact velocity is increased: 
At low velocities the crack structure is determined by the interference 
of elastic waves resulting in an essentially two-dimensional crack pattern with a regular 
structure. High velocity impact gradually excites cracking in the 3D 
bulk of the solid giving rise to a highly disordered crack structure 
and a steeper decay of the mass distribution.
In 3D bulk samples the transition is hindered so that a unique exponent 
emerges. In spite of the observed non-universality of the complete mass distribution, 
identifying subsets of fragments dominated by different cracking mechanisms
an astonishing universality of their mass distributions is revealed at all 
plate thicknesses. Our study provides a robust 
scenario which leads to the energy dependence of mass distribution exponents in fragmentation
phenomena but it still underlines the importance of universality.

\section{Discrete element model of fragmentation}
We investigate the fragmentation of plate-like objects induced by 
impact of a projectile in the framework of a discrete element model (DEM)
developed recently \cite{PhysRevE.88.062207}. 
The model has proven successful in reproducing key features of fracture processes
of heterogeneous materials. 
Here we briefly summarize the main steps of the model construction based on Ref.\ \cite{PhysRevE.88.062207}.
Similar modelling approaches have been also used in Refs.\ 
\cite{PhysRevLett.112.065501,carmona_fragmentation_2008,PhysRevE.86.016113}.

The sample is represented 
as a random packing of spherical particles which was generated by sedimenting
particles in a rectangular container \cite{PhysRevLett.112.065501,PhysRevE.88.062207}. 
The diameter $d$ of the particles was sampled from a uniform distribution in a 
narrow range $\left< d\right>-\Delta d/2 \leq d \leq \left< d\right>+\Delta d/2$,
where $\left< d\right>$ denotes the average diameter. The range $\Delta d$ of diameter
values was set as $\Delta d/\left< d\right>= 0.05$.
In the simulations plate-like samples were constructed with a rectangular basis of side
length $L$ and height $H$. Simulations were carried out with a fixed extension
$L=30$  varying the height of the sample $H$ in the range $H=3-15$ measured 
in units of the average particle diameter $\left<d\right>$. 
The total number of particles in the samples falls between 5000 ($H/\left<d\right>=3$)
and 25000 ($H/\left<d\right>=15$).

In the model cohesive interaction of particles is provided by beams 
which connect the particles along the edges of a Delaunay
triangulation of the initial particle positions. In 3D the total deformation of
a beam is calculated as the superposition of elongation, torsion, as well as
bending and shearing. Crack formation is captured such that the beams,
modeling cohesive forces between grains, can be broken according to a physical
breaking rule, which takes into account the stretching and bending of contacts
\begin{eqnarray}
\left(\frac{\varepsilon_{ij}}{\varepsilon_{th}}\right)^2 + \frac{\max(\Theta_i, 
\Theta_j)}{\Theta_{th}} \geq 1.
\end{eqnarray}
Here $\varepsilon_{ij}$ denotes the axial strain of the beam between particles $i$ 
and $j$, while 
$\Theta_i$, and $\Theta_j$ are the bending angles of the beam ends.
The parameters $\varepsilon_{th}$ and $\Theta_{th}$ control 
the relative importance of the two breaking modes 
\cite{kun_study_1996,carmona_fragmentation_2008,PhysRevE.86.016113,PhysRevLett.112.065501,
PhysRevE.88.062207}. 
In the model there is only structural disorder
present, i.e.\ the breaking thresholds are constant $\varepsilon_{th}=0.002$ and
$\Theta_{th}=2^o$, however, the physical properties
of beams such as length, cross section, and elastic moduli, are determined by the random 
particle packing. 
At the broken beams along the surface of the spheres cracks are
generated inside the solid and as a result of the successive beam breaking the
solid falls apart. The interaction of those particles which are not connected 
by beams, e.g.\ because the beam has been broken, is described 
by the Hertz contact law \cite{poschel_grandyn_2005}. The Hertz contact ensures
that force can be transmitted through crack faces when they are pressed against
each other.
The fragments are defined as sets of discrete particles
connected by the remaining intact beams. The time evolution of the fragmenting solid
is obtained by solving the equations of motion of the individual particles 
\cite{allen_computer_1984,poschel_grandyn_2005} until
the entire system relaxes meaning that no beam breaking occurs during one
thousand consecutive time steps and there is no energy stored in deformation.
For more details of the model construction and parameter settings 
see Refs.\ \cite{PhysRevE.88.062207}.
\begin{figure}
\begin{center}
\epsfig{bbllx=50,bblly=30,bburx=360,bbury=330,file=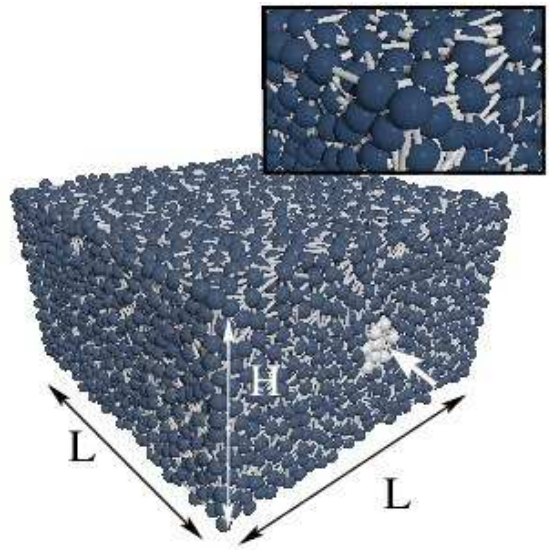,
  width=7.5cm}
\caption{(Color online)
Geometrical setup of the simulations: rectangular samples of a square shaped
basis were considered in such a way that the side length $L/\left<d\right>=30$ of the square
was fixed and the height $H/\left<d\right>$ of the sample was varied from $3$ to $15$.
Here a sample is presented for $H/\left<d\right>=11$. A particle in the middle of the front
side of the sample was selected, together with its neighbors it got an initial velocity
pointing inward the sample. 
The cylinders connecting the particles represent beams and the 
white arrow indicates the direction of the impact velocity. The inset shows
a closer view on a small segment of the sample.
}
\label{fig:illust}
\end{center}
\end{figure}

Impact loading was performed in such a way that
a single surface particle was selected in the middle of one of the side walls of the 
sample. Together with its contacting neighbors it got an initial velocity $\vec{v}_0$ 
pointing
towards the center of mass of the body. This is equivalent to an experimental 
setup 
where the impactor does not penetrate the target but it is stopped after hitting
the target surface as e.g.\ in Refs.\ \cite{kadono_fragment_1997,kadono_fragment_2005}. 
The geometrical setup of the simulations and
the loading condition of impact is illustrated in Fig.\ \ref{fig:illust}.
For the smallest plate thickness $H/\left<d\right>=3$
the impact site practically spans the cross section of the samples while for higher 
thicknesses the loading condition gets close to a point-like impact.
\begin{figure}
\begin{center}
\includegraphics[bbllx=5,bblly=0,bburx=395,bbury=330,scale=0.6]{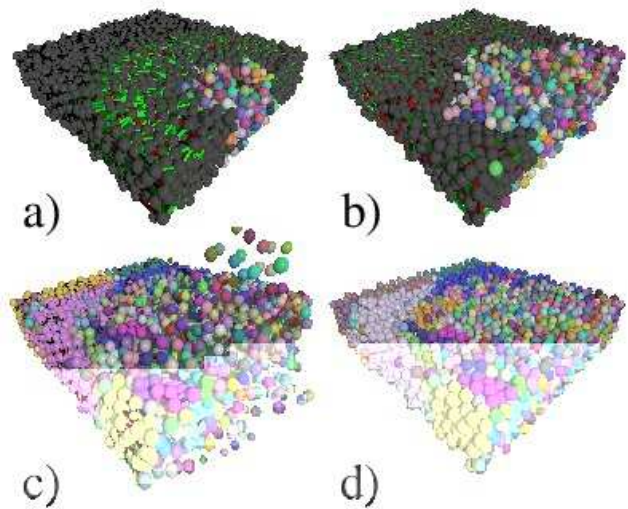}
  \caption{(Color online) Time evolution of a fragmenting plate of thickness 
  $H/\left<d\right>=3$ embedded in the three-dimensional space. The impact velocity 
  $v_0/c=0.2$ is slightly above the critical point of fragmentation.
  Compressed beams are green while the stretched ones are red 
  so that the propagation of a shock wave can be observed. 
  Particles are colored according to the fragment they belong to
  while the color (grey) of fragments is randomly selected from a palette. 
  $(d)$ presents the final state of the 
  system where the sample is reassembled by placing the particles 
  back to their original position. 
   \label{fig:crack}}
\end{center}
\end{figure}
Computer simulations were performed to determine the sound speed $c$ 
of the model material. In the presentation of the results lengths and velocities 
are made dimensionless by dividing them with the average particle diameter 
$\left<d\right>$ and with the sound speed $c$, respectively.

\section{Damage-fragmentation transition}

\begin{figure}
\begin{center}
\includegraphics[bbllx=0,bblly=0,bburx=355,bbury=305,scale=0.65]{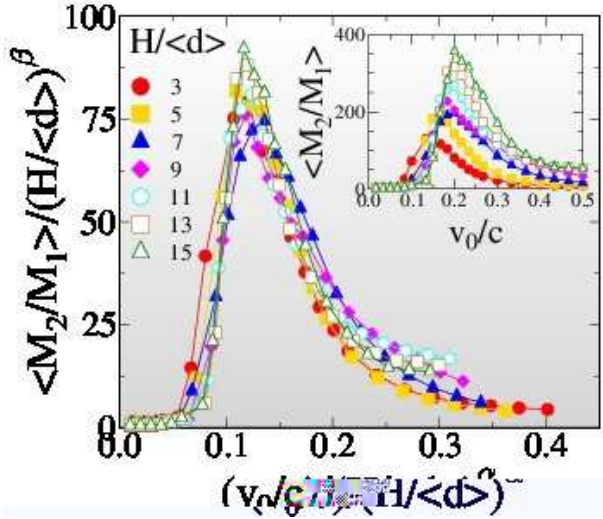}
  \caption{(Color online) Inset: Average mass of fragments $\left<M_2/M_1\right>$ as a function of 
  the impact velocity $v_0$ for 
  all plate thicknesses $H$ considered. The main panel presents the same data rescaled
  with appropriate powers of the plate thickness $H$ to obtain collapse 
  of the curves.
   \label{fig:m2_m1_all}}
\end{center}
\end{figure}

In order to investigate how the overall geometry of the system affects the 
outcomes of the breakup
process for each plate thickness $H$ we carried out simulations varying the 
impact velocity in a broad range. To accumulate statistics simulations were 
repeated 2000 times for each parameter set with different realizations 
of the structural disorder. As a representative example Fig.\ \ref{fig:crack} 
presents the time evolution of a plate of thickness $H/\left<d\right>=3$ generated by an 
impact with initial velocity $v_0/c=0.2$. The sample breaks into a 
large number of pieces
due to the shock wave generated by the impact resulting in a broad distribution 
of fragment sizes. The figure also shows the final reassembled body where fragments
can be easily identified.
Of course, the degree of breakup strongly depends 
on the value of the impact velocity $v_0$: at low $v_0$
the sample just gets damaged around the impact site, i.e.\ some beams break 
and small fragments comprising a few particles are ejected but the main part of the body
retains its integrity. 
To achieve complete 
fragmentation, where even the largest fragment is significantly smaller than the 
original body, the impact velocity has to exceed a critical value $v_c$. 
To quantify the degree of breakup 
we determined the average mass of fragments $\left<M_2/M_1\right>$ 
as the ratio of the second $M_2$ and first $M_1$ moments 
of fragment masses.
The $k$th moment $M_k$ of the fragment mass in a single simulation is defined as 
\beq{
M_k=\sum_i m_i^k -m_{max}^k,
\label{eq:momk}
}
where $m_i$ denotes the mass of single pieces, while $m_{max}$ is the largest 
fragment mass. The sum runs over all fragments.
The ratio of the two moments $M_2/M_1$ was determined in single simulations and then 
it was averaged over fragmentation events at each impact velocity $v_0$. 
The inset of Fig.\ \ref{fig:m2_m1_all} shows that 
gradually increasing $v_0$ the average fragment mass increases due to the creation of
larger fragments. Since the largest fragment is always removed from the moments in 
Eq.\ (\ref{eq:momk}),
the decreasing branch of $\left<M_2/M_1\right>$ is caused by the absence of a 
dominating piece. Hence, the position of the sharp maximum can be identified 
with the critical value $v_c$ of the impact velocity where complete breakup occurs. 
The result demonstrates 
that the system undergoes a transition as the impact velocity is varied from 
the damage phase ($v_0<v_c$) characterized by the presence of a dominating piece,
to the fragmentation phase ($v_0>v_c$) where no major fragment prevails.
The existence of the damage-fragmentation transition has been verified for 
various types of systems both by  experiments 
\cite{hayakawa_impact_1996,katsuragi_scaling_2003,
katsuragi_crossover_2004,astrom_universal_2004,katsuragi_explosive_2005,
dos_santos_experimental_2010,ji_fragmentation_2010} and computer simulations
\cite{kun_study_1996,kun_transition_1999,wittel_fragmentation_2004-1,
behera_fragmentation_2005,kun_scaling_2006,carmona_fragmentation_2008,
PhysRevE.86.016113,astrom_universality_2000,Myagkov2014120}.

In the inset of Fig.\ \ref{fig:m2_m1_all} the critical velocity $v_c$ is an increasing 
function of $H$ because the total mass of the sample $M_{tot}$ 
increases with the plate thickness  $M_{tot}\propto L^2H$.
The main panel of Fig.\ \ref{fig:m2_m1_all} presents that rescaling $v_0$ and 
$\left<M_2/M_1\right>$ of the inset
with appropriate powers of the plate thickness $H$ the results 
obtained at different thicknesses 
can be collapsed on a master curve. The high quality collapse implies the validity 
of the scaling form 
\begin{equation}
\left<M_2/M_1\right>(v_0,H)=H^{\beta}\phi(v_0/H^{\alpha}), 
\label{eq:m2m1scale}
\end{equation}
where $\phi(x)$ denotes the scaling function. The exponents were obtained 
numerically as $\alpha=0.2(3)$ and $\beta=0.5(2)$ giving the best collapse 
in Fig.\ \ref{fig:m2_m1_all}. 
It follows from Eq.\ (\ref{eq:m2m1scale}) that the critical impact velocity 
$v_c$ increases as a power law of the plate thickness
\begin{equation}
v_c \propto H^{\alpha}.
\end{equation}
In the following we focus on the probability distribution of the mass of fragments
$p(m)$ in the fragmented phase to understand how it evolves with the impact velocity
at different plate thicknesses.

\section{Mass distribution of fragments}
\begin{figure}
\begin{center}
\includegraphics[bbllx=40,bblly=10,bburx=750,bbury=640,scale=0.34]{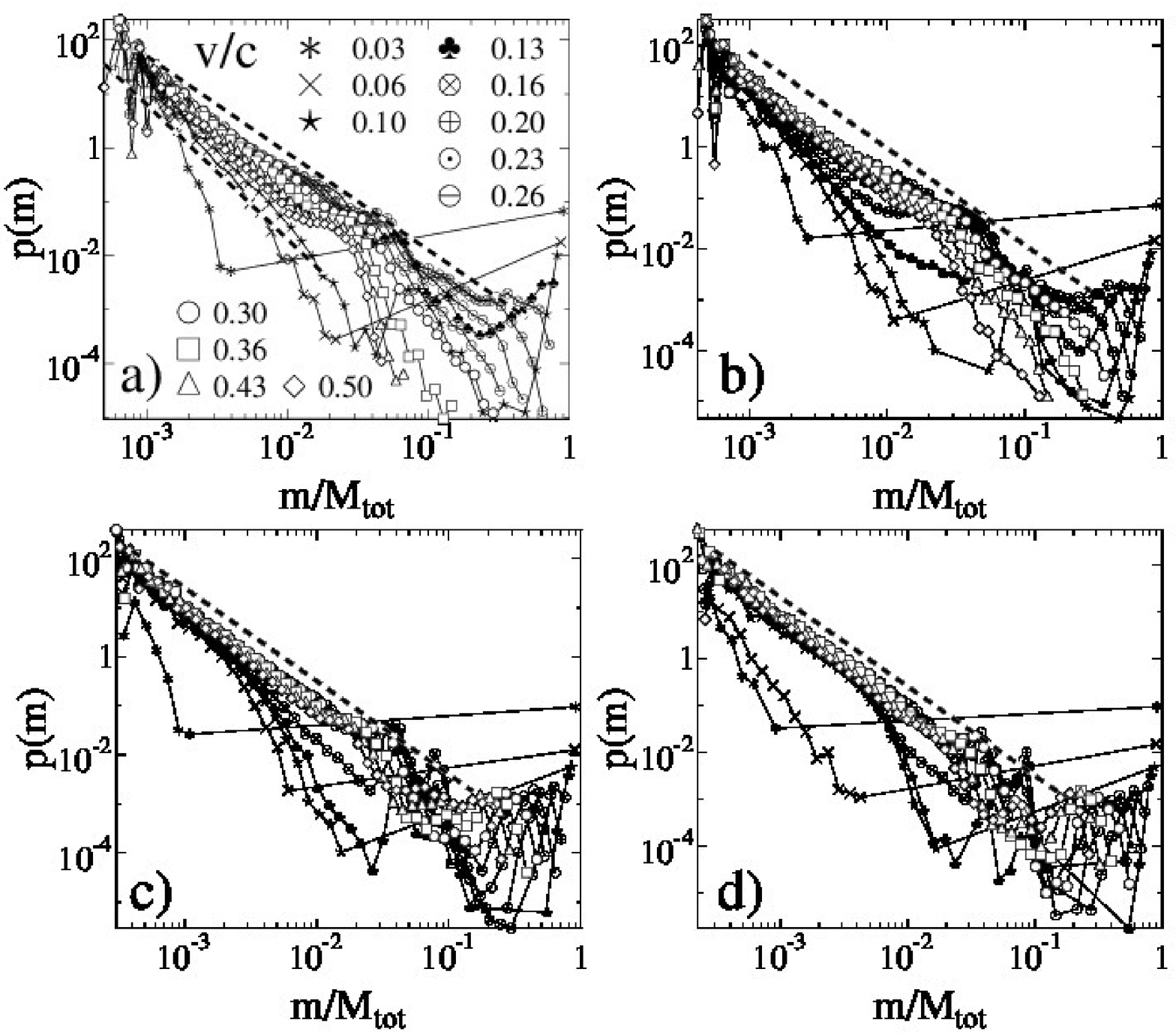}
  \caption{Mass distribution of fragments at different impact velocities for 
four different plate
  thicknesses $H/\left<d\right>$: $(a)$ $3$, $(b)$ $5$, $(c)$ $11$, $(d)$ $15$. 
  In the damage phase the distributions are composed 
of two distinct parts, i.e.\ the large residues form a peak at $m/M_{tot} \approx 1$, 
while the small fragments have a rapidly decreasing distribution. 
The two regimes are separated by a gap which gradually 
disappears as the critical velocity $v_c$ is approached from below.
The dashed straight lines represent power laws of exponents $(a)$ 1.7 and 2.4, 
$(b)$ 1.7, $(c)$ 1.9, and $(d)$ 1.9.
   \label{fig:massdist_4}}
\end{center}
\end{figure}

Figure \ref{fig:massdist_4} presents the fragment mass distribution 
$p(m)$ for several values of the plate 
thickness $H$ at different impact velocities. 
A generic feature of the distributions is that in the
damage phase ($v_0< v_c$) $p(m)$ is composed of two distinct parts: due to the 
presence of a big residue, a peak of the mass distribution 
is formed close to $m/M_{tot} \approx 1$ while the 
distribution of small pieces has a rapidly decreasing functional form. The two 
regimes are
separated by a gap which gradually disappears as the critical impact velocity 
is approached 
from below. It can be observed that in the fragmented phase  ($v_0 > v_c$) 
small sized fragments have a power law mass distribution 
\begin{equation}
p(m) \propto m^{-\tau},
\end{equation}
which is followed by a cutoff regime. The power law over a broad range 
first occurs at the critical point.

The important feature of our results presented in Fig.\ \ref{fig:massdist_4}
is that for low plate thicknesses $H$ we observe a gradual
increase of the mass distribution exponent from $\tau=1.7$ obtained at the 
critical point $v_c$ to $\tau=2.4$ reached in the limit of high $v_0$ values 
(see Fig.\ \ref{fig:massdist_4}$(a)$). However, increasing the plate thickness 
in Figs.\ \ref{fig:massdist_4}$(b,c,d)$ this dependence of $\tau$ on the impact velocity
gradually disappears and for high
plate thicknesses $H/\left<d\right>\geq 11$ only a single value of the exponent  $\tau=1.9$ remains. 
For increasing $v_0$ the mass of the largest fragment must decrease 
which may result in an apparent increase of the exponent simply due to the 
shifting cutoff of the distributions. However, contrary to Ref.\ \cite{PhysRevE.86.016113} 
our analysis showed that 
the change of $\tau$ in Fig.\ \ref{fig:massdist_4} is the real behavior 
of $p(m)$, it cannot be transformed 
out by rescaling with the average fragment mass. 

\subsection{Fragmentation mechanisms}
Our computer simulations revealed that the observed dependence of $\tau$ on the 
impact
velocity is caused by a transition between two distinct fragmentation 
mechanisms, 
which emerges due to the interplay of the geometry of the sample and of the 
dimensionality 
of the embedding space.
It can be seen in Fig.\ \ref{fig:crack} that immediately after impact the 
specimen 
gets damaged in the vicinity of the impact site, 
i.e.\ in a small volume starting from the surface all beams get broken and 
single particles (powder in the model)
are ejected from the specimen. 
The impact loading generates a shock wave which gets gradually attenuated
by the breaking of beams and by the expansion over a larger volume. 
At sufficiently high $v_0$ the compression wave can pass through the sample 
and reflects back with opposite phase as a tensile wave at the free boundaries 
of the rectangular specimen freely evolving in the three dimensional embedding 
space.
In this velocity range the stress field which gives rise to the formation 
of extended cracks and final breakup is determined by the interference 
pattern of compression and reflected tensile waves.  
\begin{figure}
\begin{center}
\includegraphics[bbllx=45,bblly=25,bburx=520,bbury=450,scale=0.50]{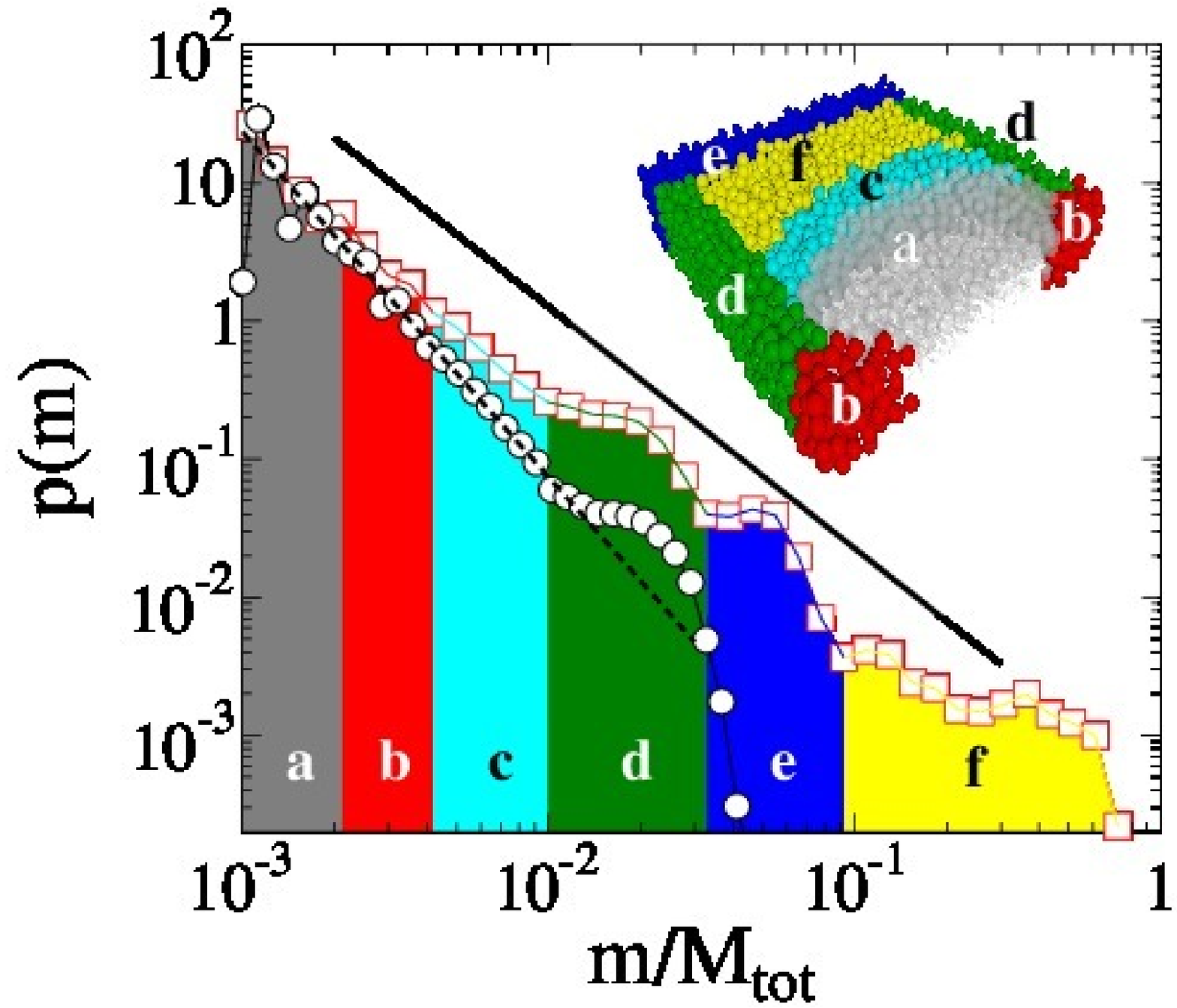}
  \caption{(Color online)
  Mass distribution of fragments $p(m)$ obtained at the critical point of fragmentation 
  $v_c$ for the smallest 
  thickness $H/\left<d\right>=3$. Fragments giving dominating contribution to different 
  ranges of $p(m)$ originate from well defined spatial regions 
  of the sample. These regions are highlighted in a single sample in the inset
  using the same colors (characters) as for the corresponding ranges of the mass distribution.
  For comparison we also present the mass distribution at the highest 
  impact velocity $v_0/c=0.5$ where the power law regime is significantly steeper.
  The slopes of the two straight lines are $\tau=1.7$ and $\tau=2.4$.
   \label{fig:mass_maxim}}
\end{center}
\end{figure}
This mechanism has the consequence that for the limit of thin plates $H\ll L$ 
the breakup of the specimen at the critical impact velocity is caused by a
relatively regular crack pattern which is essentially two-dimensional. 
To demonstrate how this fragmentation mechanism works Fig.\ \ref{fig:mass_maxim}
presents the mass distribution $p(m)$ of pieces at the critical impact velocity 
for $H/\left< d\right>=3$ together with a sample in the inset where particles 
of different spatial regions are highlighted by different colors. 
It can be observed that
$p(m)$ has a power law behavior over a broad range of fragment masses,
however, it is decorated by distinct maxima. Detailed analysis revealed that
the emergence of the maxima is the fingerprint of the regularity of the 
two-dimensional crack structure of the plate-like object, i.e.\ 
fragments giving dominating contribution to 
a maximum always emerge in the same spatial region of the specimen.
To make it clear in Fig.\ \ref{fig:mass_maxim} we assigned colors (characters) to the maxima 
of $p(m)$ such that in the inset the same colors (characters) are used for the particles of the 
spatial regions where the corresponding fragments originate from. 
It can be seen that the smallest fragments comprising a few particles ($grey$ (a)) 
are generated in the destructed
zone close to the impact point, while larger fragments are formed deeper in the sample 
($cyan$ (c)). The interference of elastic waves 
generates a highly stretched
zone along the surface of the plate which gives rise to the detachment of 
surface fragments 
both on the left and right sides ($green$ (d)) and on the back side ($blue$ (e)). 
The two corners of the front side of the sample ($red$ (b)) result in a slight 
local maximum of $p(m)$ between the $grey$ (a) and $cyan$ (c) regions. 
The largest fragment ($yellow$ (f)) controlling the cutoff of the distribution is 
created inside 
the specimen close to the back side with a shape elongated perpendicular to the 
direction of impact. The two maxima of the yellow region of $p(m)$ are caused by 
the fact that the yellow region breaks into two major pieces with a high 
probability. 

As the impact velocity gets high enough 
the overall two dimensional character of the crack pattern disappears and most 
of the 
cracks are created in the three-dimensional bulk of the material. 
This second fragmentation mechanism 
gradually becomes dominating with increasing $v_0$. As a consequence, in Fig.\ 
\ref{fig:massdist_4}$(a)$ the cutoff of $p(m)$
shifts toward smaller $m$ and the fraction of large fragments decreases in the 
mass 
distribution giving rise to a higher value of the power law exponent $\tau$. 
It follows from the above arguments that the dependence of the mass distribution
exponent $\tau$ on the impact velocity $v_0$ is caused by
the gradual crossover from the planar two-dimensional to the  
three-dimensional bulk crack structure. At the critical velocity
the 2D character dominates while in the limit of high $v_0$ the crack pattern 
is 
completely three-dimensional. In the intermediate velocity range both mechanisms
are present so that the observed mass distribution is a blend of their 
contributions.
The crossover is gradual in the sense that in the highly destroyed zone around the 
impact site 
the crack pattern is three-dimensional already at the lowest impact velocities 
which then
spreads over the sample as $v_0$ increases. To have a clear view on the two 
limits of $p(m)$
with different exponents of the power law regimes in Fig.\ \ref{fig:mass_maxim} 
the mass
distribution is also presented for the highest impact velocity $v_0/c=0.5$ we 
considered.
The two straight lines of the figure represent power laws of exponents 
$\tau=1.7$ and 
$\tau=2.4$.

\subsection{Superposition of subsets of fragments}
The changing exponent of the mass distribution is the consequence of the gradual
crossover of the crack structure in the sample as the impact velocity increases.
The two- and three-dimensional crack structures favor fragment formation in 
different 
spatial regions of the sample having also different extensions. 
Hence, in order to understand how the crossover emerges 
we identified sets of fragments according to their position in the sample
and analyzed how their 
contributions to the complete mass distributions evolve with the impact velocity 
and plate thickness. 
The key feature of fragments is whether they are created 
in the bulk or on the surface of the initial body. 
Since the sample surface is rather irregular, to identify
the position of fragments we construct their bounding box and compare the 
location of its
corner points to the bounding box of the original sample. For this purpose in the final 
state 
of the fragmentation process we reassembled the sample, 
as it is illustrated in Fig.\ \ref{fig:crack}$(d)$,
and determined the bounding box of each fragment aligning the sides of the box
with the edges of the sample. 
\begin{figure}
\begin{center}
\epsfig{bbllx=0,bblly=0,bburx=250,bbury=345,
file=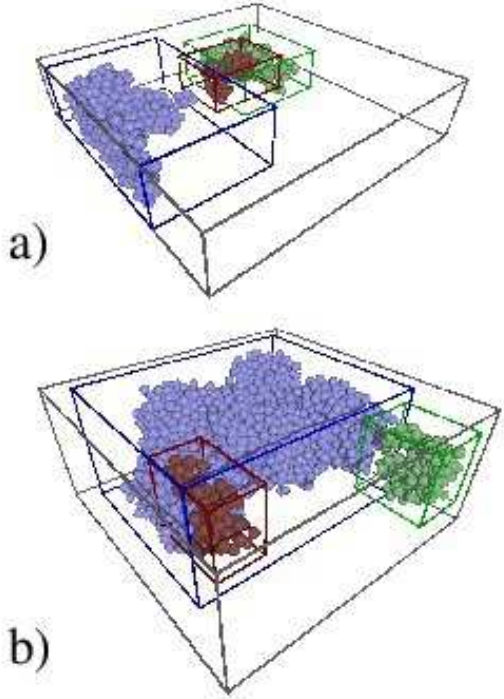,
  width=7.0cm}
\caption{(Color online)
Identification of fragment subsets based on the bounding box of fragments and of
the complete sample. Two plates are shown with different thicknesses and impact
velocities $(a)$ $H/\left<d\right>=5$ and $v_0/c=0.23$, $(b)$ 
$H/\left<d\right>=11$ 
and $v_0/c=0.3$. For each subset the largest fragment is highlighted with 
different colors: light blue (light grey), green (medium grey), and red (dark grey) 
stand for the spanning, surface, and bulk 
fragments, respectively. The bounding boxes are indicated by the wire frames.
}
\label{fig:bbox}
\end{center}
\end{figure}
Based on the position and extension of the bounding box three 
types of fragments are distinguished:
\begin{itemize}
  \item {\it Bulk fragment}: if the corners of the bounding box all lie inside 
        the sample, i.e.\ their distance from the surface of the bounding box of 
        the sample is greater than a threshold distance $0.2\left<d\right>$, 
        the fragment is considered to be a bulk fragment.

  \item {\it Surface fragment}: if any of the corners, but not all of them, 
        are within the threshold distance to the surface of the sample's 
        bounding box, the fragment is called surface fragment.
        
  \item {\it Spanning fragment}: those fragments which span the sample at least 
  in one direction, are called spanning fragments. For spanning fragments 
  all the corners of the bounding box lie in the vicinity of the sample surface.
\end{itemize}
Figure \ref{fig:bbox} demonstrates the identification of the three subsets of 
fragments 
in two plates of different thicknesses. For each subset a single fragment is 
highlighted
together with its bounding box.

\begin{figure}
\begin{center}
\epsfig{bbllx=25,bblly=10,bburx=740,bbury=620,file=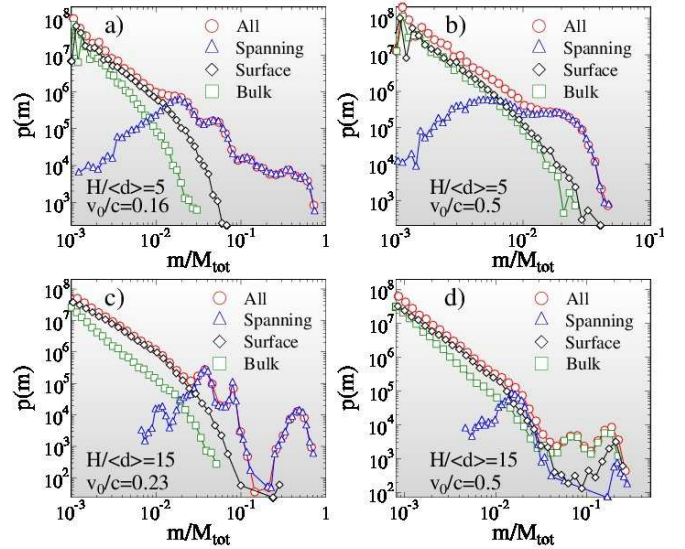,
  width=8.7cm}
\caption{(Color online)
Mass distribution of subsets of surface, spanning, and bulk fragments together 
with the complete distribution of all fragments for two plate thicknesses $H/\left<d\right>=
5$ and $15$ in the upper and lower rows, respectively. 
In both cases the results are presented for two values of the impact velocity 
$v_0$ slightly above the corresponding $v_c$ (left column) and for the 
limit of high speed impact (right column). }
\label{fig:subsets_massdist}
\end{center}
\end{figure}

The spanning fragments are typically formed by cracks, which connect two 
opposite sides
of the sample. Such cracks emerge due to 
the global interference pattern of elastic waves.
In thin plates below and at the critical impact velocity  $v_c$
most of the mass is comprised in spanning fragments; bulk and surface pieces 
can mainly be formed around the destroyed zone at the impact site. 
Above $v_c$ the fraction of spanning fragments rapidly decreases, however, they 
always have the largest mass 
so that the spanning fragments determine the cutoff of the mass distribution 
$p(m)$ at any $v_0$. 
Figure \ref{fig:subsets_massdist} presents the mass distribution of the three
subsets of fragments for a plate of thickness $H/\left<d\right>=5$ 
at the critical impact velocity together with the complete distribution. 
Note that the partial distributions are normalized such that their integral is equal to 
their fraction in the complete set of fragments. The 
Figure clearly demonstrates that the cutoff and the large mass regime of $p(m)$ is
controlled by fragments which span the sample in the direction perpendicular to 
the plate.

In thin plates surface fragments are generated from spanning fragments by cracks 
formed in the 
bulk of the specimen segmenting the spanning cracks. Due to this geometric 
constraint, in Fig.\ \ref{fig:subsets_massdist}$(a)$ the low mass regime 
of $p(m)$ is dominated by surface fragments, while bulk pieces have 
only a minor contribution. It follows that for thin plates the overall power law 
character 
of the complete mass distribution $p(m)$ originates from a mainly two dimensional
crack pattern. The power law regime of the complete mass distribution covers a 
broad range
of fragment masses due to the spanning fragments and their daughter pieces 
on the surface of the sample. The exponent $\tau$ of the power law regime
was obtained by fitting $\tau=1.7\pm 0.05$ (compare also to Fig.\ 
\ref{fig:massdist_4}).

Fig.\ \ref{fig:subsets_massdist}$(b)$ presents the same mass distributions 
at the limit of high impact velocities.
Here the range of the power law regime of the complete mass distribution $p(m)$ 
gets reduced and the exponent
increases to a high value $\tau=2.4\pm 0.07$ (compare also to Fig.\ 
\ref{fig:massdist_4}). 
As the impact velocity increases bulk cracking 
gets activated which leads to 
a three-dimensional crack pattern with a highly disordered structure. 
Consequently, the difference of surface 
and bulk fragments disappears, they both have the same mass distribution 
with a high exponent in Fig.\ \ref{fig:subsets_massdist}$(b)$. Still the 
spanning fragments have the largest mass 
but they are only the remainings of the
detached pieces. 

Fig.\ \ref{fig:subsets_massdist}$(c,d)$ present the corresponding results for 3D 
bulk bodies
with $H/\left<d\right>=15$.
Compared to the plate-like objects of Fig.\ \ref{fig:subsets_massdist}$(a,b)$ 
one
can observe that the spanning pieces do not have a dominant role, at low 
velocities they are
formed by detachment, while at high velocities they are just the corners of the 
sample.
Spanning fragments give rise to distinct humps of the distribution at high $m$ 
values, however, 
also surface fragments have contribution to the cutoff of $p(m)$. Fig. 
\ref{fig:massdist_4}
demonstrated that for bulk bodies the fragment mass distribution exponent practically 
does not depend on the impact velocity, it has a unique value $\tau=1.9\pm 0.05$. 
According to Figs.\ \ref{fig:subsets_massdist}$(c,d)$
the reason of the constant exponent is that the relative 
fraction of surface and bulk pieces does not depend on the impact velocity.

\subsection{Universality of partial mass distributions}
\begin{figure}
\begin{center}
\epsfig{bbllx=15,bblly=5,bburx=730,bbury=615,file=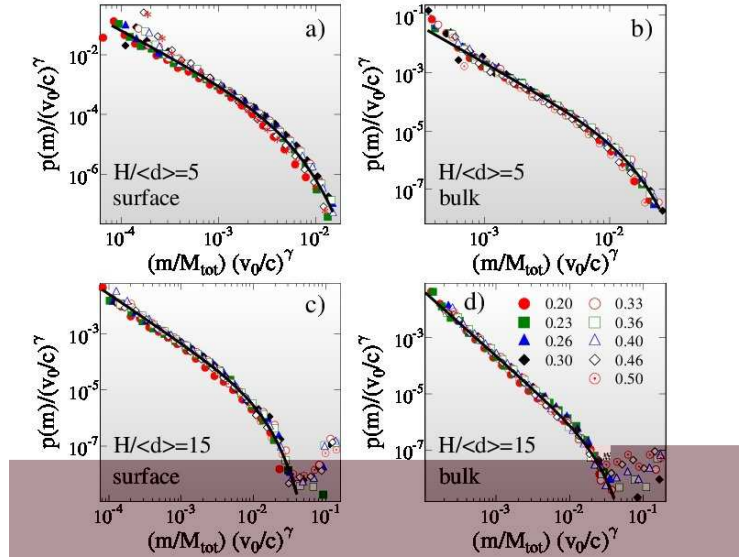,
  width=8.7cm}
\caption{(Color online)
Data collapse analysis of the mass distribution of subsets of surface
and bulk fragments for two plate thicknesses $H/\left<d\right>=$
5 and 15 in the upper and lower rows, respectively. 
Rescaling the distributions with appropriate powers of the impact velocity
above the critical point good quality data collapse is achieved. The bold lines 
represent fits of the master curves with Eq.\ (\ref{eq:psi}).}
\label{fig:subsets_massdist_scaling}
\end{center}
\end{figure}

A very interesting outcome of our study is that in spite of the velocity 
dependence of the exponent observed for the 
complete mass distribution, the partial distributions 
of the subsets of bulk and surface fragments exhibit a high degree of 
universality. Figure \ref{fig:subsets_massdist_scaling} presents the scaling plot
of the mass distributions of surface and bulk fragments obtained at different 
impact velocities for two plate thicknesses. It can be observed that 
rescaling the distributions with appropriate powers of the impact 
velocity good quality data collapse is obtained in all cases. 
This scaling analysis demonstrates that the partial distributions obey the 
scaling law 
\beq{
p(m)=v_0^{\gamma}\Psi(mv_0^{\gamma}),
}
where the exponent $\gamma$ depends on the plate thickness $H$.
Note that due to the normalization of the distributions the same
exponent $\gamma$ has to be used along the horizontal and vertical axis.
In Fig.\ \ref{fig:subsets_massdist_scaling} best collapse was obtained 
with the exponents $H/\left<d\right>=5$:
$\gamma=1.3$ (surface) and $\gamma=0.2$ (bulk); $H/\left<d\right>=15$:
$\gamma=0.7$ (surface) and $\gamma=0.25$ (bulk).
The scaling function $\Psi(x)$ was fitted with the functional form
\beq{
\Psi(x) \propto x^{-\tau}\exp{\left(-\left(x/x^*\right)^{\kappa}\right)},
\label{eq:psi}
}
where the exponent $\kappa$ and the characteristic scale $x^*$ only control 
the shape of the cutoff. 
The most remarkable feature of the results is that best fits of the scaling 
function $\Psi$  is obtained with $\tau=1.7$ and $\tau=2.4$ for 
surface and bulk fragments
respectively, for all thicknesses. This result implies that the partial 
distributions exhibit universality as it has been observed for a broad class 
of fragmentation phenomena. The observed non-universality of the complete 
distribution of all fragments originates from the blending of the distributions of
subsets of fragments whose contributions depend both on the impact velocity and on 
the plate thickness.

\section{Discussion}
We investigated the impact induced breakup of heterogeneous brittle materials
in the framework of a three-dimensional discrete element model focusing on the 
mass distribution of fragments. Based on large scale computer simulations
we resolved recent debates on the universality of the power law 
exponent
of the mass distribution which is crucial both from scientific point
of view and for industrial applications.
Simulations were carried out to investigate the impact induced breakup 
of plate-like objects where both the thickness of the plate and the impact velocity 
were varied in a broad range.
Our computer simulations revealed that for thin plates embedded in the 
three-dimensional 
space the power law exponent of the fragment mass distribution  has a strong 
dependence on the impact velocity:
power law is first obtained at the critical velocity of impact with an 
exponent $\tau=1.7$ which then gradually increases to $\tau=2.4$ for 
high impact velocities. However, for 3D bulk samples a 
unique exponent is obtained $\tau=1.9$, dependence 
on the impact velocity can only be pointed out for the cutoff of the 
distributions. 
Note that the value of the exponent $\tau=1.7$
of $p(m)$ falls close to the theoretical prediction of Ref.\ 
\cite{astroem_universal_2004,kekalainen_solution_2007}
based on the branching-merging scenario of dynamic cracks:
if fragments are formed by the merging of branches of splitting unstable cracks 
a universal exponent of the fragment mass distribution $\tau=(2D-1)/D$ 
was predicted depending solely on the dimensionality $D$ of the embedding space. 
For $D=3$ the formula yields $\tau=5/3$ in the vicinity of our numerical 
result, although, in our case simulations did not reveal a branching-merging
sequence of cracks. The exponent $\tau=1.9$ of 3D bulk samples is consistent with other
DEM results, e.g.\ the same exponent was obtained for the fragmentation 
of brittle spheres impacted against a hard wall 
\cite{carmona_fragmentation_2008,PhysRevE.86.016113}. 

The reason of the velocity dependent exponent 
is that due to the interplay of the 
geometry of the sample and of the dimensionality of the embedding space a crossover
occurs between two different fragmentation mechanisms. In the vicinity of the 
critical impact velocity the crack pattern is essentially two-dimensional 
determined by the interference 
pattern of compressive and tensile waves generated by the impact. 
This crack pattern has a high degree of regularity which gives rise to local 
maxima of the fragment mass distribution
on the overall power law functional form. 
At increasing impact velocities 
bulk cracking gets activated so that the crack structure becomes 
three-dimensional with a high degree of randomness. 

A similar effect of the interference pattern
of elastic waves has been observed for slender rods  where 
fragmentation was induced by a hit at the free rod end. The mass 
distribution of pieces proved to have discrete humps at certain fractions of the
buckling wave length \cite{PhysRevLett.94.035503} similar to what 
we obtained for plates. Studying the impact induced 
breakup of thin glass plates, in the experiments of Ref.\ \cite{kadono_crack_2002}
an increase of the mass distribution exponent was reported with 
increasing impact velocity. The authors argued that the effect can be attributed 
to the increase of the fractal dimension of the crack pattern, i.e.\ as the 
crack structure gets more-and-more space filling the mass distribution exponent
increases and approaches a limit value \cite{kadono_crack_2002}. Our results 
clarify the background of these experimental findings unveiling the underlying 
mechanism. 

Comparing the bounding box of fragments and that of the complete sample 
we decomposed the fragment ensemble into subsets of bulk, surface, and 
spanning pieces. The formation of these fragments is governed by different 
cracking mechanisms. Scaling analysis showed a striking universality of
the mass distributions of bulk and surface fragments with strongly different 
exponents. The results imply that the velocity dependence 
of the exponent of the complete mass distribution at intermediate
velocities $v_0$ is observed due to the mixing of the contributions of the subsets 
of fragments, where the mixing 
ratio depends on $v_0$. Our results have the general consequence 
that energy dependence of the mass distribution exponent of fragmentation 
phenomena can be expected when a low dimensional object is embedded 
into a higher dimensional space allowing for the emergence of a transition 
in the spatial structure of cracks generated by the initial shock wave.

\begin{acknowledgments}
The work is supported by the projects No.\ TAMOP-4.2.2.A-11/1/KONV-2012-0036.
The projects are implemented through the New Hungary Development Plan,
co-financed by the European Union, European Social Fund and the European
Regional Development Fund. We acknowledge the support of OTKA K84157.
\end{acknowledgments}

\bibliography{/home/feri/papers/statphys_fracture}

\end{document}